\documentstyle[12pt,a4wide]{article}

\input epsf

\begin{document}

%%% name: mymacros.tex

\newcommand{\kiste}[1] {\centering { \leavevmode #1}}

\newcommand{\abbildung}[2]
        {
        \begin{figure}[ht]
        \centering
        \kiste
                {
                \epsfbox{#1}
                }
        #2
        \end{figure}
        }

%%%%%%%%%%%%%%%%%%%%%%%%%%%%%%%%%%%%%%%%%%%%%%%%%%%%%%%%%%%%%%%%%%%%%%%%%%%

\renewcommand{\abbildung}[2] { }

%%%%%%%%%%%%%%%%%%%%%%%%%%%%%%%%%%%%%%%%%%%%%%%%%%%%%%%%%%%%%%%%%%%%%%%%%%%

\newcommand{\beq} {\begin{equation}}
\newcommand{\eeq} {\end{equation}}
\newcommand{\bea} {\begin{eqnarray}}
\newcommand{\eea} {\end{eqnarray}}

\newcommand{\RR}{{\rm I \!\!\, R}}
\newcommand{\ZZ}{{\bf Z}}

\newcommand{\connplus}[2] { {\rm conn_+} (#1,#2) }
\newcommand{\connminus}[2] { {\rm conn_-} (#1,#2) }
\newcommand{\conngroe}[2] { {\rm conn_>} (#1,#2) }
\newcommand{\connklei}[2] { {\rm conn_<} (#1,#2) }

\newcommand{\connplusinv}[2] { {\rm conn_+^{-1}} (#1,#2) }
\newcommand{\connminusinv}[2] { {\rm conn_-^{-1}} (#1,#2) }
\newcommand{\conngroeinv}[2] { {\rm conn_>^{-1}} (#1,#2) }
\newcommand{\connkleiinv}[2] { {\rm conn_<^{-1}} (#1,#2) }

\newcommand{\crossing}[1] { {\cal C}(#1) }
\newcommand{\crossings}[2] { {\cal C}(#1,#2) }

\newcommand{\bild} {{\rm image}}
\newcommand{\perm} {{\rm perm}}
\newcommand{\tr} {{\rm tr}}
\newcommand{\mod} {{\,\, \rm mod \,\,}}
\newcommand{\sgn} {{\rm sgn}}
\newcommand{\linknum} {{\rm lk}}

\newcommand{\calK} { {\cal K} }

\newcommand{\topKreis} {I_{/\sim}}
\newcommand{\projFlaeche} {\RR^2}
\newcommand{\abbilden} {\rightarrow}
\newcommand{\kreuzungen} { {\cal C} }
\newcommand{\indexmenge} { {\cal I} }
\newcommand{\darausFolgt} { \Rightarrow }
\newcommand{\anderenfalls} {\mbox{ otherwise }}

\newcommand{\esExistiert} { \exists \,}
\newcommand{\fuerAlle} { \bigwedge }
\newcommand{\ggIntervall}[2] { [ #1,#2 ] }
\newcommand{\goIntervall}[2] { [ #1,#2 [ \,\, }
\newcommand{\ogIntervall}[2] { \,\, ] #1,#2 ] }
\newcommand{\ooIntervall}[2] { \,\, ] #1,#2 [ \,\, }
\newcommand{\klartext}[1]  {\,\,\, {\rm #1} \,\,\,}

\newcommand{\gleichungNr}[1] { (\ref{#1}) }

\newcommand{\zweiZeilenSumme}[2]
 { \sum_  {  #1   \atop  #2  } }

\renewcommand{\theequation}{\arabic{section}.\arabic{equation}}

%%% name: abstract.tex
\hfill DO-TH 95/02
\newline
\newline
\newline
\newline
\newline
\begin{center}
{\Large {Derivation of the total twist from }}\\
\vspace{4mm}
{\Large {Chern-Simons theory}} \\
\vspace{4cm}
{\large
	{
        Allen C. Hirshfeld and
        Uwe Sassenberg\footnote
                {
                e-mail: sassen@het.physik.uni-dortmund.de.
                Supported by the {\it Deutsche Forschungsgemeinschaft}.
                }
        }
} \\
\vspace{0.8cm}
{\it {Institut fr Physik}}\\
{\it {Universitt Dortmund}}\\
{\it {44221 Dortmund, Germany}}\\
\vspace{4cm}
{\bf{Abstract}}
\end{center}
\noindent
The total twist number, which represents the first non-trivial
Vassiliev knot invariant, is derived
from the second order expression of the Wilson loop expectation
value in the Chern-Simons theory.
Using the well-known fact that the analytical expression is an invariant,
a non-recursive formulation of the
total twist based on the evaluation of knot diagrams is constructed
by an appropriate deformation of the knot line in the
three-dimensional Euclidian space. The relation to the original definition
of the total twist is elucidated.
\newline
\newline
\pagebreak
\section{Introduction}
\setcounter{equation}{0}
The use of Chern-Simons theory as a tool for knot theory has proved
to be very fruitful.
Some of the background of the relationship of these two areas is
recalled very briefly in order to fix the starting point for the present
work.

One considers the Chern-Simons quantum field theory on the manifold $S^3$
with some gauge group, say $SU(N)$. It is characterized by a generating
functional
\beq
Z[J] = \int {\cal D} A\,\, \exp
\left\{
	i S_{\rm CS}[A] + \int_{S^3} d^3x\,\, J^{\mu}_a(x) A_{\mu}^a(x)
\right\},
\eeq
where $S_{\rm CS}$ is the integral of the Chern-Simons differential form:
\beq
S_{\rm CS}[A] = \frac k {4\pi} \int_{S^3} \tr
\left\{
	A\wedge dA + \frac 2 3 A\wedge A\wedge A
\right\}
\eeq
and $A=A_\mu^a T_a dx^\mu$ is a gauge field in some
appropriately normalized representation $T$.
Here $k$ must be an integer in order to maintain gauge invariance.
The observables of the theory are Wilson-line operators along framed links.
The concept of framing was introduced in the original work of Witten
\cite{Witten}.
The Wilson loops are symbolically written as path-ordered exponentials of
line integrals along knots:
\beq
W(K) = \tr P \exp
\left\{
	\oint_K A
\right\}
\mbox{\hspace{1cm}for a knot $K$ embedded in $S^3$},
\nonumber
\eeq
\beq
W(L) = W(K_1) \ldots W(K_n)
\mbox{\hspace{1cm}for a link with $n$ components $L = \{K_1,\ldots,K_n\}$}.
\eeq
In the framework of quantum field theory the expectation values of $W(L)$
are defined as
\beq
\langle W(L) \rangle_f =
\frac
	{
	\int {\cal D} A\,\, W(L) \exp \{ i S_{\rm CS}[A] \}
	}
	{
	\int {\cal D} A\,\, \exp \{ i S_{\rm CS}[A] \}
	},
\eeq
where $f$ symbolizes a framing prescription used in the evaluation of
the Wilson loop.
The expectation values are dependent on the coupling constant $k$
and on the order $N$ of the gauge group (and on the framing $f$).
It has been shown that
$\langle W(L) \rangle$ associated to a link $L$ fulfils the skein relation
of a so-called generalized Jones polynomial.
A comprehensive exposition of this subject can be found in
a monograph by E. Guadagnini \cite{Guadagnini}.

The Wilson loops can be expanded in a perturbation series.
Then $\langle W(L) \rangle$ is a power series in
$1/k$ and the Casimir factors of the gauge group
$c_2(T)$, $c_v$, $c_4(T)$ etc.
Its coefficients are multi-dimensional path-ordered line integrals
along the knot. Since $1/k$ and the Casimir factors can be considered as
linearly independent the coefficient of every monomial is a link invariant.
These line integrals can be calculated either numerically or by expanding
the generalized Jones polynomial mentioned above, as demonstrated
in a recent paper by Alvarez and Labastida \cite{AlLa}.

We present here a direct method which does not rely on recursive
procedures.
Our starting point is the analytical expression for
$\langle W(K) \rangle$ in the second order of $1/k$ which has been
calculated by Guadagnini et al. in \cite{GuMaMi}:
\beq
\langle W(K) \rangle_{f, \rm 2. order} =
\dim T
\left(
	\frac {2\pi} {k}
\right)^2
\left\{
	-\frac 1 2 \,
	c_2^2(T) \varphi_f^2(K)
	+
	c_v c_2(T) [\rho_1(K)+\rho_2(K)]
\right\},
\eeq
where $\varphi_f(K)$ is the framing number of the framed knot.
The line integrals $\rho_1(K)$ and $\rho_2(K)$ are given by
\bea
\rho_1(K) & = &
-\frac 1 {32\pi^3}
    \int_{K} dx_1^{\mu_1}
    \int_{\rm BP}^{x_1} dx_2^{\mu_2}
    \int_{\rm BP}^{x_2} dx_3^{\mu_3} \,\,
\epsilon^{\nu_1 \nu_2 \nu_3}
\epsilon_{\mu_1 \nu_1 \sigma_1}
\epsilon_{\mu_2 \nu_2 \sigma_2}
\epsilon_{\mu_3 \nu_3 \sigma_3} \nonumber\\
& \times &
\int d^3z \frac {(z-x_1)^{\sigma_1}} {|z-x_1|^3}
          \frac {(z-x_2)^{\sigma_2}} {|z-x_2|^3}
          \frac {(z-x_3)^{\sigma_3}} {|z-x_3|^3} \nonumber \\
\eea
\bea
\rho_2(K) & = & \frac 1 {8\pi^2}
 \int_{K} dx_1^{\mu_1} \int_{\rm BP}^{x_1} dx_2^{\mu_2}
 \int_{\rm BP}^{x_2} dx_3^{\mu_3} \int_{\rm BP}^{x_3} dx_4^{\mu_4} \,\,\,
 \epsilon_{\mu_4 \mu_2 \sigma_2} \epsilon_{\mu_3 \mu_1 \sigma_1} \nonumber \\
	& \times & \frac {(x_4 - x_2)^{\sigma_2}} {|x_4-x_2|^3}
              \frac {(x_3 - x_1)^{\sigma_1}} {|x_3-x_1|^3} . \nonumber \\
\eea
They are frame-independent, as shown in \cite{GuMaMi}.
The knot invariant we are interested in is
\beq
\rho^{\rm II}(K) := \rho_1(K) + \rho_2(K).
\eeq
Our method to calculate $\rho^{\rm II}$ is based
on the evaluation of knot diagrams $\calK$ only, without recurse
to skein relations.
The program can be summed up in the following steps.
\begin{itemize}
\item[-]
We are interested of course in the evaluation of $\rho^{\rm II}$ for
arbitrary knots embedded in three-dimensional space.
To achieve this it is convenient to work with the knot diagrams.
The formalism appropriate for our purposes is developed
in sections \ref{definitionOfTheKnotDiagrams} to
\ref{sectionCrossingNumbers}.
\item[-]
The sense in which knot diagrams are flattened knots is clarified in
section \ref{flatKnotLimit}.
\item[-]
The integral $\rho_2$ is calculated for the limit of flat knots. The
result is an expression for $\rho_2$ based on the evaluation of
knot diagrams (section \ref{berechnungRhoZwei}).
\item[-]
A similar expression for $\rho_1$ is constructed using some
properties of the line integral $\rho_1(K)$ {\sl and the invariance of
$\rho^{\rm II}(K)$}. The knot theoretical formulation of
$\rho^{\rm II}(\calK)$ is thereby completed
(section \ref{berechnungRhoEins}).
\item[-]
It is known from \cite {GuMaMi}
that $\rho^{\rm II}(K)$ is closely related to the total
twist $\tau$ defined by Lickorish and Millet in \cite{LiMi}.
This relation is examined for our diagrammatical version of $\rho^{\rm II}$
(section \ref{relationRhoIITotalTwisting}).
\item[-]
The procedure for calculating the invariants $\rho^{\rm II}$ and $\tau$
for a given knot using our expressions \gleichungNr{mainResult} and
\gleichungNr{totalTwist}, which are based on the crossing numbers
introduced in section \ref{crossingNumbers}, is illustrated in
section \ref{beispiel}.
\end{itemize}
We should emphasize that the important point in this paper is not just
to find another expression for the total twist, but to find a method
of constructing it which can be generalized to other link
invariants.
As has been argued in \cite{AlLa} the line integrals from the Wilson loop
expansion are Vassiliev invariants, if normalized correctly.
These have been introduced by Vassiliev in \cite{Va}.
For an axiomatic approach to this topic see
Birman and Lin \cite{BiLi}.
The total twist, refered to above, is the simplest non-trivial invariant
of this kind \cite{Birman}.
The method presented here, extended to higher orders,
should allow interesting insights concerning this large class of
knot invariants.
%
%
%
%%% name: diagdefs.tex
\section{Definition of the knot diagrams}
\setcounter{equation}{0}
\label{definitionOfTheKnotDiagrams}
In this section we shall give the definitions of the knot diagrams used in
this paper.
Based, oriented knots are considered.
Their diagrams will be described as mappings from a circle to a
plane. First, define the unit interval $I=[0,1]$ and an equivalence relation
which identifies $0 \sim 1$. Then $\topKreis$ is homeomorphic to a circle.
The projection plane is $\projFlaeche$.
The shadow diagram of some knot is given by a mapping
\beq
\pi : \topKreis \abbilden \projFlaeche .
\eeq
The point $\pi(0) = \pi(1)$ is called the basepoint.
The diagram itself is
\beq
K = \bild (\pi) \subset \projFlaeche .
\eeq
The set of crossings is given by
\beq
\kreuzungen = \{ x \in \RR^2 \mid \esExistiert {t_1 \not= t_2 \in \topKreis}:
\pi(t_1) = \pi(t_2) = x \} \subset \RR^2 .
\eeq
The over/under crossing information is given by a function
\beq
\epsilon : \kreuzungen \abbilden \{-1,+1\},
\eeq
defined as in the accompanying figure.
%%%%%%%%%%%%%%%%%%%%%%%%%%%%%%%%%%%
\abbildung{kreuzung.eps}{}
%%%%%%%%%%%%%%%%%%%%%%%%%%%%%%%%%%%%%%%%%%%%%%%
Then $\pi^{-1}(\kreuzungen)$ is a set $\{s_0,\ldots,s_{2n-1}\}$,
with $s_i \in \topKreis$ and $n$ the number of crossings.
More generally, we set $\pi^{-1}(\kreuzungen) =
\{s_i\mid i\in \indexmenge \}$,
where $\indexmenge$ is some index set.
The mapping $\pi$ may be chosen such that for all $i\in \indexmenge$
$s_i \not\sim 0$, i.e. the basepoint does not coincide with any crossing.
The knot diagram is now defined by
\beq
\cal K = (\pi,\epsilon,\indexmenge).
\eeq
The order relations $>$ and $<$ in $I$ induce relations $>$ and $<$ in the
index set $\indexmenge$ according to
\beq
\label{ordnungDerIndexmenge}
s_i > s_j \darausFolgt i > j,\,\,\,\, i,j\in\indexmenge.
\eeq
If not denoted otherwise $\indexmenge = \{0,\ldots,2n-1\}$ with its common
order will be used. In this case the addition (modulo $2n$) of elements of
$\indexmenge$ with integers is defined and will be used.
On $\topKreis$ intervals will be denoted as follows.
For $s_i \not\sim s_j$, $\ggIntervall{s_i}{s_j} $
is the closed arc from $s_i$ to $s_j$ following the orientation of
$\topKreis$.
Hence
\beq
\ggIntervall{s_i}{s_j} \cup \ggIntervall{s_j}{s_i} = \topKreis
\klartext{and}
\ggIntervall{s_i}{s_j} \cap \ggIntervall{s_j}{s_i} = \{ s_i,s_j \}.
\eeq
{}From this, {\sl ordered} subsets of the index set are defined, and denoted as
\beq
\ggIntervall{i_a}{i_b} = \{ i\in \indexmenge \mid
 s_i\in \ggIntervall{s_{i_a}}{s_{i_b}} \}
 \mbox{\,\,\,for\,\,\,} i_a \not= i_b ,
\eeq
with the property
\beq
\ggIntervall{i_a}{i_b} \cup \ggIntervall{i_b}{i_a} = \indexmenge
\klartext{and}
\ggIntervall{i_a}{i_b} \cap \ggIntervall{i_b}{i_a} = \{ i_a,i_b \}.
\eeq
We define {\sl open} subsets of the index set by
\beq
\ooIntervall{i_a}{i_b} = \ggIntervall{i_a}{i_b} \setminus \{i_a,i_b\} .
\eeq
The following alternative notations for the crossing function will also be
used:
\beq
\begin{array}{rcl}
\klartext{For \,\,\, all}
i_1\in\indexmenge
&
\klartext{define}
&
\epsilon(i_1) := \epsilon(\pi(s_{i_1}))\\
\klartext{For \,\,\, all}
i_1,i_2\in\indexmenge
&
\klartext{define}
&
\epsilon(i_1,i_2) :=
\left\{
\begin{array}{l}
\epsilon(i_1) \klartext{if} \pi(s_{i_1}) = \pi(s_{i_2})
\klartext{and} i_1\not=i_2 \\
0 \anderenfalls \\
\end{array}
\right. \\
\end{array}
\eeq
An example for a knot diagram ${\cal K} = (\pi,\epsilon,\indexmenge)$
defined in this way is shown in figure \ref{achter}.
Here, $n=4$ (figure-eight knot) and $\indexmenge = \{ 0,1,\ldots,7 \}$.
%%%%%%%%%%%%%%%%%%%%%%%%%%%
\abbildung{achter.eps}{\caption{Diagram of the figure-eight knot.}
                       \label{achter} }
%%%%%%%%%%%%%%%%%%%%%%%%%
\paragraph{Definition of pieces of the knot diagram.}
We consider a knot diagram $\calK=(\pi,\epsilon,\indexmenge)$.
A piece of the knot diagram will be defined as an open connection between
two crossings or one crossing with itself.
Consider two indices $i_1 \not= i_2\in\indexmenge$.
Then a set such as
\beq
S = \pi(\ooIntervall{s_{i_1}}{s_{i_2}}) \subset \RR^2
\eeq
is called a piece of the knot diagram. To this piece an index subset
is associated:
\beq
\indexmenge(S) = \ooIntervall{i_1}{i_2} = \{i_1+1,i_1+2,\ldots,i_2-1\}.
\eeq
Of course, a piece is not necessarily free of self-crossings.

Two pieces $S$ and $T$ will be called non-overlapping if
$\indexmenge(S) \cap \indexmenge(T) = \emptyset$. In this case
$S$ and $T$ can intersect each other, but the set of common points
only consists of crossings.
%
%
%
%%% name: reidemei.tex
\section{Description of the Reidemeister moves}
\setcounter{equation}{0}
It is necessary for the following to formulate the Reidemeister moves
in terms of $\pi$, $\epsilon$, and $\indexmenge$.
In this section $\calK=(\pi,\epsilon,\indexmenge)$ will always denote
the the knot diagram before the Reidemeister move, and
$\calK'=(\pi',\epsilon',\indexmenge')$ will be the diagram afterwards.
The sets of crossings are denoted by $\kreuzungen$ and $\kreuzungen'$.
The Reidemeister moves can be formulated in the following way.
In the diagram $L_{\rm I}^{+/-}$, the indices which form the crossing are
$k_1$ and $k_1+1$, and the situation can obviously be described by
\beq
\pi(s_{k_1}) = \pi(s_{k_1+1}).
\eeq
After performing the Reidemeister-I move (see diagram $L_{\rm I}^{0}$
in figure
(\ref{reidemeisterI})),
i.e. replacing $\pi$ by a mapping $\pi'$,
the situation is described by a new index set $\indexmenge'$
and a new crossing function $\epsilon':\kreuzungen'\abbilden \{-1,+1\}$,
where $\kreuzungen' = \kreuzungen\setminus\{\pi(s_k)\}$.
\beq
\indexmenge'=\indexmenge \setminus \{ k_1,k_1+1 \} ,\,\,\,
\epsilon'= \epsilon\mid _{\kreuzungen '}.
\eeq
%%%%%%%%%%%%%%%%%%%%%%%%%%%%%%%%%%%
\abbildung{l_i.eps}{\caption{Reidemeister-I move.}
                    \label{reidemeisterI}    }
%%%%%%%%%%%%%%%%%%%%%%%%%%%%%%%%%%%%%%%%%%%%%%%

For the second Reidemeister move two cases have to be distinguished.
First we consider the situation $L_{\rm II-A}$ in figure
\ref{reidemeisterIIA}
with the indices $k_1$, $k_1+1$, $k_2$, and $k_2+1$.
It is characterized by
$$
\pi(s_{k_1}) = \pi(s_{k_2}) ,\,\,\,
\pi(s_{k_1+1}) = \pi(s_{k_2+1}) ,\,\,\,
$$
\beq
\epsilon(k_1,k_2) = -\epsilon(k_1+1,k_2+1).
\eeq
In the situation $L_{\rm II-A}^0$, i.e. after the Reidemeister-II-A move,
we have
\beq
\indexmenge' = \indexmenge \setminus \{ k_1,k_1+1,k_2,k_2+1 \} ,\,\,\
\epsilon'= \epsilon \mid_{\kreuzungen '}.
\eeq
%%%%%%%%%%%%%%%%%%%%%%%
\abbildung{l_ii_a.eps}{\caption{Reidemeister-II-A move.}
                       \label{reidemeisterIIA} }
%%%%%%%%%%%%%%%%%%%%%%%%%%%%%%%%%%%%%%%%%%%%%%%

For $L_{\rm II-B}$ (see figure \ref{reidemeisterIIB}) with the same
indices as in $L_{\rm II-A}$ we have
$$
\pi(s_{k_1}) = \pi(s_{k_2+1}),\,\,\,
\pi(s_{k_2}) = \pi(s_{k_1+1}),\,\,\,
$$
\beq
\epsilon(k_1,k_2+1) = -\epsilon(k_1+1,k_2).
\label{epsilonRelationRMII}
\eeq
Performing the Reidemeister-II-B move we get the same $\indexmenge'$ and
$\epsilon'$ as for the Reidemeister-II-A move.
%%%%%%%%%%%%%%%%%%%%%%%
\abbildung{l_ii_b.eps}{caption{Reidemeister-II-B move.}
\label{reidemeisterIIB} }
%%%%%%%%%%%%%%%%%%%%%%%%%%%%%%%%%%%%%%%%%%%%%%%

A Reidemeister-III situation as shown in figure
(\ref{reidemeisterIII}) with indices
$k_1$, $k_1+1$,
$k_2$, $k_2+1$,
$k_3$, and $k_3+1$, and a
cyclic orientation is described by
$$
\pi(s_{k_1}) = \pi(s_{k_3+1}),\,\,\,\,
\pi(s_{k_2}) = \pi(s_{k_1+1}),\,\,\,\,
\pi(s_{k_3}) = \pi(s_{k_2+1})
$$
\beq
\label{zyklischeQuadratischeEpsilonRelation}
\epsilon(k_1)\epsilon(k_2)+
\epsilon(k_2)\epsilon(k_3)+
\epsilon(k_3)\epsilon(k_1)= -1.
\eeq
The Reidemeister-III move does not change the index set, i.e.
$\indexmenge' = \indexmenge$. The relation between the crossing
functions $\epsilon$ and $\epsilon'$ is:
\beq
\label{rmDreiEpsilonEpsilonStrich}
\epsilon'(k_1) = \epsilon(k_2),\,\,\,\,
\epsilon'(k_2) = \epsilon(k_3),\,\,\,\,
\epsilon'(k_3) = \epsilon(k_1).
\eeq
The relation \gleichungNr{zyklischeQuadratischeEpsilonRelation}
is also valid for $\epsilon'$.
%%%%%%%%%%%%%%%%%%%%%%%
\abbildung{l_iii.eps}{\caption{Reidemeister-III move.}
                      \label{reidemeisterIII}  }
%%%%%%%%%%%%%%%%%%%%%%%%%%%%%%%%%%%%%%%%%%%%%%%
%
%
%
%%% name: kreuzahl.tex
\section{Crossing numbers}
\setcounter{equation}{0}
\label{sectionCrossingNumbers}
\subsection{Definitions}
\label{crossingNumbers}
We shall now define some functions of knot diagrams.
Later they will be used to formulate the invariant $\rho^{\rm II}$ for
diagrams.
Let $\calK = (\pi,\epsilon,\indexmenge)$ be a knot diagram,
with the set of crossings $\kreuzungen$ and the number of crossings $n$.
The behaviour of $n$ under the Reidemeister moves can be
easily found by counting the crossings. It is
\bea
n(L_{\rm I}^{+/-}) - n(L_{\rm I}^0)  & = & 1 \nonumber\\
n(L_{\rm II-A/B}) - n(L_{\rm II-A/B}^0)  & = & 2  \nonumber\\
n(L_{\rm III}^+) - n(L_{\rm III}^-) & = & 0.\nonumber\\
\eea
The writhe number, which is a regular isotopy invariant, is defined
by
\beq
\chi_1(\calK) = \sum_{c\in\kreuzungen} \epsilon(c).
\eeq
This can also be written as
\beq
\chi_1(\calK) =
 \zweiZeilenSumme{ j_1,j_2\in\indexmenge } { j_1>j_2 }
 \epsilon(j_1,j_2).
\eeq
For two non-overlapping pieces $S$ and $T$ of the diagram the first
crossing number is defined as
\beq
\chi_1(S,T) = \sum_{i_S\in\indexmenge(S)} \sum_{i_T\in\indexmenge(T)}
\epsilon(i_S,i_T).
\eeq
Instead of writing $\chi_1(S,T)$ the notation using index sets
$\chi_1(\indexmenge(S),\indexmenge(T)):=\chi_1(S,T)$ will also be used.
If $S$ and $T$ start and end at the same point $c$
(see figure \ref{linkPart}),
$\chi_1(S,T)$ is closely related to
the linking number $\lambda$ of the two link components arising from
{\sl nullifying} $c$. Denoting these components as $L_S$ and $L_T$ one
has
\beq
\frac 1 2 \chi_1(S,T) = \lambda(L_S,L_S).
\label{relationLinkingChiEins}
\eeq
Finally, we define an object called the {\sl second self-crossing
number} which is {\sl not} an invariant:
\beq
\chi_2(\calK) =
 \zweiZeilenSumme{ j_1,j_2,j_3,j_4\in\indexmenge }{ j_1>j_2>j_3>j_4 }
\epsilon(j_1,j_3) \epsilon(j_2,j_4).
\eeq
It can easily be shown that $\chi_2$ can also be written as
\beq
\label{chiZweiSymmetrischeDarstellung}
\chi_2(\calK) = \frac 1 4 \sum_{ j_1,j_3\in\indexmenge }
 \epsilon(j_1,j_3) \chi_1(\ooIntervall{j_1}{j_3},\ooIntervall{j_3}{j_1}).
\eeq
In this last formulation neither the basepoint nor the order of the
elements of $\indexmenge$, which was fixed by the basepoint, appear.
Therefore $\chi_2(\calK)$ is independent of the
basepoint.
Now some properties of $\chi_2(\calK)$ will be examined.
%%%%%%%%%%%%%%%%%%%%%%%%%%%
\abbildung{linkpart.eps}{\caption{Two link components and two pieces of a
 crossing.} \label{linkPart} }
%%%%%%%%%%%%%%%%%%%%%%%%%%
%
%
%
%
\subsection{Behaviour of $\chi_2$ under change of one crossing}
\label{einKreuzungsVerhalten}
An important property of $\chi_2$, which will be used later, is its
behaviour if the sign of one crossing is changed.
Consider two knot diagrams $\calK = (\pi,\epsilon,\indexmenge)$
and $\calK' = (\pi,\epsilon',\indexmenge)$ which differ only in one crossing
$c$, i.e.
\beq
\label{relationEpsilonEpsilonStrich}
\epsilon'(d) =
\left\{
\begin{array}{l}
	-\epsilon(d) \klartext{if} d=c \\
        +\epsilon(d) \klartext{if} d\not=c. \\
\end{array}
\right.
\eeq
The crossing is formed by two indices $i_c^+,i_c^- \in \indexmenge$.
We want to calculate
$\chi_2(\calK) - \chi_2(\calK')$. It is convenient to use the representation
(\ref{chiZweiSymmetrischeDarstellung}).
The range of summation $\indexmenge$ is divided into
$\{i_c^+,i_c^-\}$ and $\indexmenge \setminus \{i_c^+,i_c^-\}$.
Then
\bea
\chi_2(\calK) - \chi_2(\calK')
& = &
\frac 1 2 \epsilon(i_c^+,i_c^-)\chi_1
	(\ooIntervall{i_c^+}{i_c^-} , \ooIntervall{i_c^-}{i_c^+} ; \epsilon)
\nonumber \\
& + &
\frac 1 4 \sum_{j_1,j_3\in\indexmenge\setminus\{i_c^+,i_c^-\}}
	\epsilon(j_1,j_3)\chi_1
	(\ooIntervall{j_1}{j_3} , \ooIntervall{j_3}{j_1} ; \epsilon)
\nonumber \\
& - &
\frac 1 2 \epsilon'(i_c^+,i_c^-)\chi_1
	(\ooIntervall{i_c^+}{i_c^-} , \ooIntervall{i_c^-}{i_c^+} ; \epsilon')
\nonumber \\
& - &
\frac 1 4 \sum_{j_1,j_3\in\indexmenge\setminus\{i_c^+,i_c^-\}}
	\epsilon'(j_1,j_3)\chi_1
	(\ooIntervall{j_1}{j_3} , \ooIntervall{j_3}{j_1} ; \epsilon') ,
\nonumber \\
\eea
where the additional argument of $\chi_1$ indicates which crossing function
is used.
The first and the third summands give
\beq
\label{ersterUndDritterSummand}
\epsilon(i_c^+,i_c^-)\chi_1
	(\ooIntervall{i_c^+}{i_c^-} , \ooIntervall{i_c^-}{i_c^+} ; \epsilon).
\eeq
In the second and fourth terms the $\epsilon$-part can be factorized,
since $\epsilon(j_1,j_3) = \epsilon'(j_1,j_3)$
in this subset of $\indexmenge$. The remaining factor is the difference
of the $\chi_1$-parts:
\bea
& & \hspace{-2cm}
\chi_1 (\ooIntervall{j_1}{j_3} , \ooIntervall{j_3}{j_1} ; \epsilon)
-
\chi_1 (\ooIntervall{j_1}{j_3} , \ooIntervall{j_3}{j_1} ; \epsilon')
\nonumber \\
& = &
\left\{
\begin{array}{l}
	\begin{array}{r}
		2\epsilon(i_c^+,i_c^-)
		\klartext{if}
		i_c^+\in\ooIntervall{j_1}{j_3}
		\klartext{and}
		i_c^-\in\ooIntervall{j_3}{j_1} \\
		\klartext{or}
		i_c^-\in\ooIntervall{j_1}{j_3}
		\klartext{and}
		i_c^+\in\ooIntervall{j_3}{j_1} \\
		\end{array} \\
	\begin{array}{r}
                0 \anderenfalls . \\
	\end{array} \\
\end{array}
\right.
\nonumber \\
\eea
This follows immediately from the definition of $\chi_1$ and the
relation (\ref{relationEpsilonEpsilonStrich}).
The condition
\beq
	i_c^+\in\ooIntervall{j_1}{j_3}
	\klartext{and}
	i_c^-\in\ooIntervall{j_3}{j_1}
\eeq
is equivalent to the condition
\beq
	j_1\in\ooIntervall{i_c^-}{i_c^+}
	\klartext{and}
	j_3\in\ooIntervall{i_c^+}{i_c^-} .
\eeq
Therefore the second and the fourth terms give
\bea
\label{zweiterUndVierterSummand}
[
	\sum_{ j_1\in\ooIntervall{i_c^+}{i_c^-} \atop
	       j_3\in\ooIntervall{i_c^-}{i_c^+} }
	& + &
	\sum_{ j_1\in\ooIntervall{i_c^-}{i_c^+} \atop
	       j_3\in\ooIntervall{i_c^+}{i_c^-} }
]
\, \frac 1 4 \epsilon(j_1,j_3) 2 \epsilon(i_c^+,i_c^-)
\nonumber \\
& = &
\chi_1( \ooIntervall{i_c^+}{i_c^-},\ooIntervall{i_c^-}{i_c^+} ;\epsilon )
\epsilon(i_c^+,i_c^-) .
\nonumber \\
\eea
Combining (\ref{ersterUndDritterSummand}) und
(\ref{zweiterUndVierterSummand}) we get
\beq
\chi_2(\calK) - \chi_2(\calK') =
2 \epsilon(i_c^+,i_c^-)
\chi_1( \ooIntervall{i_c^+}{i_c^-},\ooIntervall{i_c^-}{i_c^+} ;\epsilon ).
\eeq
\subsection{Behaviour of $\chi_2$ under the Reidemeister moves}
\label{verhaltenUnterRM}
In the following the behaviour of $\chi_2$ under the moves
RM-I, RM-II-A, RM-II-B, and RM-III will be examined.
It is not necessary to calculate the behaviour for other versions of
RM-III, e.g. with reversed orientations for some lines, because these
moves can be composed from the moves mentioned above.
\paragraph{Behaviour of $\chi_2$ under RM-I.}
Consider two knot diagrams $\calK = (\pi,\epsilon,\indexmenge)$ and
$\calK' = (\pi',\epsilon',\indexmenge')$ which are equal up to one
Reidemeister-I move, so that $\calK$ contains a situation $L_{\rm I}^{+/-}$
at the indices $k,k+1\in\indexmenge$ and $\calK'$ contains instead
a situation
$L_{\rm I}^0$, as shown in figure \ref{reidemeisterI}.
This means that $\indexmenge' = \indexmenge \setminus
\{k,k+1\}$. The basepoint is assumed to be outside the Reidemeister
situation,
i.e. somewhere in $\pi(\ooIntervall{s_{k+1}}{s_{k}})$, so that
$k+1 \mod 2n > k$.
We now calculate $\chi_2(\calK) - \chi_2(\calK')$.
There are two sums over four summation variables, each of which satisfies
$j_1>j_2>j_3>j_4$.
Since the sole difference between $\calK$ and $\calK'$ is the crossing
at $k$ and $k+1$,
all cases in which none of the summation variables is $k$ or $k+1$ cancel
out.
Consider the summands with $j_1=k$. Due to the factor $\epsilon(j_1,j_3)$
the summands can only be non-vanishing if $j_3=k+1$ which is not possible
because the summation variables fulfil the condition $j_1>j_3$.
Consider now the summands with $j_1=k+1$.
Due to the factor $\epsilon(j_1,j_3)$ there can be a contribution only
if $j_3=k$. But in this case the condition $j_1>j_2>j_3$
cannot be fulfilled and hence there are no non-vanishing summands.
The same argument is valid for all other cases.
Therefore, the behaviour of $\chi_2$ under the first Reidemeister move is
\beq
\chi_2(\calK) - \chi_2(\calK') = 0.
\eeq
\paragraph{Behaviour of $\chi_2$ under RM-II-A and RM-II-B.}
Consider a knot $\calK = (\pi,\epsilon,\indexmenge)$ which contains a
situation $L_{\rm II-A}$ with the indices $k_1$, $k_1+1$, $k_2$, and $k_2+1$,
where $k_2 +1 > k_1$.
Consider another knot $\calK' = (\pi',\epsilon',\indexmenge')$ which is
equal to $\calK$ except for containing $L_{\rm II-A}^0$ instead of
$L_{\rm II-A}$,
as shown in figure \ref{reidemeisterIIA}.
We wish to calculate $\chi_2(\calK) - \chi_2(\calK')$.
Among all summands it is sufficient to consider those with at least one
summation variable in $\{k_1,k_1+1,k_2,k_2+1\}$.
However, if in the product $\epsilon(j_1,j_3)\epsilon(j_2,j_4)$
one of the arguments, e.g. $j_1$ belongs to $\{k_1,k_1+1,k_2,k_2+1\}$,
the product is non-zero if and only if $j_3$ assumes the corresponding
value so that $j_1$ and $j_3$ form a crossing.
This reduces the number of summands which have to be considered.
\bea
\chi_2(\calK)-\chi_2(\calK')
 & = &
 \zweiZeilenSumme{ j_2,j_4\in\indexmenge }{ k_2>j_2>k_1+1, k_1>j_4 }
 \epsilon(k_2,k_1)\epsilon(j_2,j_4) \nonumber \\
 & + &
 \zweiZeilenSumme{ j_2,j_4\in\indexmenge }{ k_2>j_2>k_1+1, k_1>j_4 }
 \epsilon(k_2+1,k_1+1)\epsilon(j_2,j_4) \nonumber \\
 & + &
 \zweiZeilenSumme{ j_1,j_3\in\indexmenge }{ j_1>k_2+1, k_2>j_3>k_1+1 }
 \epsilon(j_1,j_3)\epsilon(k_2,k_1) \nonumber \\
 & + &
 \zweiZeilenSumme{ j_1,j_3\in\indexmenge }{ j_1>k_2+1, k_2>j_3>k_1+1 }
 \epsilon(j_1,j_3)\epsilon(k_2+1,k_1+1) \nonumber \\
 & + &
 \epsilon(k_2+1,k_1+1)\epsilon(k_2,k_1) \nonumber \\
\eea
Using the relation \gleichungNr{epsilonRelationRMII} the first four terms
cancel pairwise, and we are left with
\beq
\chi_2(\calK)-\chi_2(\calK')
  =  \epsilon(k_2+1,k_1+1)\epsilon(k_2,k_1)
  = -1.
\eeq
For the Reidemeister-II-B move the calculation is similar, except that
the only non-vanishing term is missing. Hence in this case
\beq
\chi_2(\calK)-\chi_2(\calK') = 0.
\eeq
%
%\bea
%\chi_2(\calK)-\chi_2(\calK')
% & = &
% \zweiZeilenSumme{ j_2,j_4\in\indexmenge }{ k_2>j_2>k_1+1, k_1>j_4 }
% \epsilon(k_2,k_1+1)\epsilon(j_2,j_4) \nonumber \\
% & + &
% \zweiZeilenSumme{ j_2,j_4\in\indexmenge }{ k_2>j_2>k_1+1, k_1>j_4 }
% \epsilon(k_2+1,k_1)\epsilon(j_2,j_4) \nonumber \\
% & + &
% \zweiZeilenSumme{ j_1,j_3\in\indexmenge }{ j_1>k_2+1, k_2>j_3>k_1+1 }
% \epsilon(j_1,j_3)\epsilon(k_2,k_1+1) \nonumber \\
% & + &
% \zweiZeilenSumme{ j_1,j_3\in\indexmenge }{ j_1>k_2+1, k_2>j_3>k_1+1 }
% \epsilon(j_1,j_3)\epsilon(k_2+1,k_1) \nonumber \\
% & = &
% \zweiZeilenSumme{ j_2,j_4\in\indexmenge }{ k_2>j_2>k_1+1, k_1>j_4 }
% \underbrace
%  {
%  \left[ \epsilon(k_2,k_1+1) + \epsilon(k_2+1,k_1) \right]
%  }_0
% \epsilon(j_2,j_4) \nonumber \\
% & + &
% \zweiZeilenSumme{ j_1,j_3\in\indexmenge }{ j_1>k_2+1, k_2>j_3>k_1+1 }
% \epsilon(j_1,j_3)
% \underbrace
%  {
%  \left[ \epsilon(k_2,k_1+1) + \epsilon(k_2+1,k_1) \right]
%  }_0
% \nonumber \\
% & = & 0
%\eea
%
\paragraph{Behaviour of $\chi_2$ under RM-III.}
The calculation of the behaviour of $\chi_2$ under the third Reidemeister
move
is longer, but in principle no more complicated than for the preceding
cases.
As in the previous cases we consider two knots:
$\calK=(\pi,\epsilon,\indexmenge)$ with a situation $L_{\rm III}^+$
consisting
of the indices $\{k_1,k_1+1,k_2,k_2+1,k_3,k_3+1\}$ and
$\calK' = (\pi',\epsilon',\indexmenge')$
with a situation $L_{\rm III}^-$ at the same location,
as shown in figure \ref{reidemeisterIII}.
The case $k_3>k_2>k_1$ is considered.
For simplicity we shall now assume\footnote{This could have been
done for the Reidemeister-II move as well.
We renounced this in order to illustrate which terms can appear in sums
of this type.} that the basepoint lies between
the index $k_3+1$ and the next index $k_3+2 \mod 2n$.
In this situation there are no summands with variables $j_i>k_3+1$
because $k_3+1$ is the greatest index in $\indexmenge$.
Only summands with two or four summation variables in
$\{k_1,k_1+1,k_2,k_2+1,k_3,k_3+1\}$ can contribute.
\bea
\chi_2(\calK) & - & \chi_2(\calK') \nonumber \\
 & = &
 \zweiZeilenSumme{j_2,j_4\in\indexmenge}{k_2>j_2>k_1+1,k_1>j_4}
 \epsilon(k_2,k_1+1)\epsilon(j_2,j_4)
 +
 \zweiZeilenSumme{j_2,j_4\in\indexmenge}{k_3>j_2>k_2+1,k_1>j_4}
 \epsilon(k_3,k_2+1)\epsilon(j_2,j_4) \nonumber \\
 & + &
 \zweiZeilenSumme{j_2,j_4\in\indexmenge}{k_3>j_2>k_2+1,k_2>j_4>k_1+1}
 \epsilon(k_3,k_2+1)\epsilon(j_2,j_4)
 +
 \zweiZeilenSumme{j_2,j_4\in\indexmenge}{k_2>j_2>k_1+1,k_1>j_4}
 \epsilon(k_3+1,k_1)\epsilon(j_2,j_4) \nonumber \\
 & + &
 \zweiZeilenSumme{j_2,j_4\in\indexmenge}{k_3>j_2>k_2+1,k_1>j_4}
 \epsilon(k_3+1,k_1)\epsilon(j_2,j_4)
 +
 \zweiZeilenSumme{j_1,j_3\in\indexmenge}{k_3>j_1>k_2+1,k_2>j_3>k_1+1}
 \epsilon(j_1,j_3)\epsilon(k_2,k_1+1) \nonumber \\
 \vspace{0.5cm}
 & - &
 \zweiZeilenSumme{j_2,j_4\in\indexmenge}{k_2>j_2>k_1+1,k_1>j_4}
 \epsilon'(k_2+1,k_1)\epsilon'(j_2,j_4)
 -
 \zweiZeilenSumme{j_2,j_4\in\indexmenge}{k_3>j_2>k_2+1,k_1>j_4}
 \epsilon'(k_3,k_1+1)\epsilon'(j_2,j_4) \nonumber \\
 & - &
 \zweiZeilenSumme{j_2,j_4\in\indexmenge}{k_3>j_2>k_2+1,k_2>j_4>k_1+1}
 \epsilon'(k_3+1,k_2)\epsilon'(j_2,j_4)
 -
 \zweiZeilenSumme{j_2,j_4\in\indexmenge}{k_2>j_2>k_1+1,k_1>j_4}
 \epsilon'(k_3,k_1+1)\epsilon'(j_2,j_4) \nonumber \\
 & - &
 \zweiZeilenSumme{j_2,j_4\in\indexmenge}{k_3>j_2>k_2+1,k_1>j_4}
 \epsilon'(k_3+1,k_2)\epsilon'(j_2,j_4)
 -
 \zweiZeilenSumme{j_1,j_3\in\indexmenge}{k_3>j_1>k_2+1,k_2>j_3>k_1+1}
 \epsilon'(j_1,j_3)\epsilon'(k_2+1,k_1) \nonumber \\
 \vspace{0.5cm}
 & - &
 \epsilon'(k_2+1,k_1)\epsilon'(k_3+1,k_2) \nonumber \\
 & - &
 \epsilon'(k_3+1,k_2)\epsilon'(k_3,k_1+1) \nonumber \\
 & - &
 \epsilon'(k_3,k_1+1)\epsilon'(k_2+1,k_1). \nonumber \\
\eea
Using the relations \gleichungNr{rmDreiEpsilonEpsilonStrich}
between $\epsilon$ and $\epsilon'$ for the
Reidemeister-III move it is easy to see that most of these terms cancel,
and we are left with
\bea
\chi_2(\calK) - \chi_2(\calK') =
& - & \epsilon'(k_2+1,k_1)\epsilon'(k_3+1,k_2) \nonumber\\
& - & \epsilon'(k_3+1,k_2)\epsilon'(k_3,k_1+1) \nonumber\\
& - & \epsilon'(k_3,k_1+1)\epsilon'(k_2+1,k_1) = +1, \nonumber\\
\eea
due to the cyclic relation (\ref{zyklischeQuadratischeEpsilonRelation})
for $\epsilon'$.
%
%
%
%%% name: flatknot.tex
\section{The limit of flat knots}
\setcounter{equation}{0}
\label{limesFlacherKnoten}
\label{flatKnotLimit}
We shall now define the limit of flat knots. As projection space
$\RR^2 \times \{0\} = \{(x,y,0) \in \RR^3\}$ will now be used.
Consider some knot diagram $\calK = (\pi,\epsilon,\indexmenge)$; the
shadow diagram is $K=\bild (\pi)$, the set of crossings is
$\kreuzungen \subset \RR^2$. We will make some assumptions for simplicity,
but without loss of generality.
The first of them is formulated as
\beq
\left| \dot{\pi}(s) \right| = const
\klartext{for\,\,\,all}
s\in\topKreis.
\eeq
For all $c\in\kreuzungen$ let $U_c\subset\RR^2$ be a sufficiently small,
open disk with center $c$. Furthermore, we define
\beq
U = \RR^2\setminus\bigcup_{c\in\kreuzungen} U_c.
\eeq
{\sl Sufficiently small} in the previous definition means that only one
crossing is contained in every $U_c$ and the basepoint $\pi(0)=\pi(1)$
lies outside every $U_c$.
The projection space is a deformation retract of $\RR^3$ with respect
to the homotopy
\beq
H_t:\RR^3\abbilden\RR^3:(x,y,z)\mapsto(x,y,t z),\,\,\, t\in [0,1].
\eeq
Hence $H_0$ projects the whole space onto the projection space and
$H_1$ is the identity in $\RR^3$.
Now we consider a knot $K_0\subset\RR^3$ parametrized by
$\pi_0:\topKreis\abbilden\RR^3$ with the following
properties:
\begin{itemize}
\item[(i)]
$H_0\circ\pi_0 = \pi$, therefore $H_0(K_0) = K$.
\item[(ii)]
For every $s\in\topKreis$
with $\pi_0(s) \in U_c\times\RR$ the third component of $\dot\pi_0(s)$
vanishes and $\ddot\pi_0(s) = 0$, this means that the knot line is straight
in this region.
\end{itemize}
Every cylinder $U_c\times\RR$ is crossed by two straight lines.
They will be called $g_c^+$ and $g_c^-$ so that
\beq
s_+ > s_- \klartext{for\,\,\,all} s_+ \in \pi_0^{-1}(g_c^+)
\klartext{and} s_- \in \pi_0^{-1}(g_c^-).
\eeq
We introduce another condition for convenience, namely
$H_0(g_c^+) \bot H_0(g_c^-)$.
The situation is shown in figure \ref{zylinder}.
%%%%%%%%%%%%%%%%%%%%%%%%%%%
\abbildung{zylinder.eps}{\caption{The cylinder $U_c\times \RR$ of a
crossing $c$.} \label{zylinder} }
%%%%%%%%%%%%%%%%%%%%%%%%%%
Consider a diagram $\calK = (\pi,\epsilon,\indexmenge)$ and an analytically
formulated ambient isotopy invariant $f(K_0)$
which is based on the evaluation of an expression defined using the
parametrization $\pi_0$ with $K_0 = \bild(\pi_0)$ and
$\pi = H_0 \circ \pi_0$.
Then
\beq
f(K_0) = \lim_{t\abbilden 0} f(H_t(K_0)) =: f(\calK)
\eeq
since $H_t$ is an ambient isotopy for every $t\in\ogIntervall{0}{1}$.
This limit will be used for calculating the
line integral $\rho_2$.
%
%
%
%%% name: rho2.tex
\section{Calculation of $\rho_2(K_0)$ for the limit of flat knots}
\setcounter{equation}{0}
\label{berechnungRhoZwei}
In this section we calculate $\rho_2(K_0)$ for the limit
of flat knots.
For any knot diagram $\calK$ a knot $K_0$ constructed as in section
\ref{limesFlacherKnoten}
may be given.
The integral $\rho_2$ for this knot is given by
\bea
\rho_2(K_0) & = & \frac 1 {8\pi^2}
 \int_{K_0} dx_1^{\mu_1} \int_{\rm BP}^{x_1} dx_2^{\mu_2}
 \int_{\rm BP}^{x_2} dx_3^{\mu_3} \int_{\rm BP}^{x_3} dx_4^{\mu_4} \,\,\,
 \epsilon_{\mu_4 \mu_2 \sigma_2} \epsilon_{\mu_3 \mu_1 \sigma_1} \nonumber \\
	& \times & \frac {(x_4 - x_2)^{\sigma_2}} {|x_4-x_2|^3}
              \frac {(x_3 - x_1)^{\sigma_1}} {|x_3-x_1|^3}, \nonumber \\
\label{rhoZwei}
\eea
where BP denotes the basepoint.
Using a parametrization
\beq
	x: \topKreis \abbilden \RR^3
\eeq
it is written as
\bea
\rho_2(K_0) & = & \frac 1 {8\pi^2}
 \int_0^1 ds_1 \int_0^{s_1} ds_2 \int_0^{s_2} ds_3 \int_0^{s_3} ds_4 \,\,\,
 \dot{x}(s_1)^{\mu_1}
 \dot{x}(s_2)^{\mu_2}
 \dot{x}(s_3)^{\mu_3}
 \dot{x}(s_4)^{\mu_4} \nonumber \\
	& \times &
 \epsilon_{\mu_4 \mu_2 \sigma_2} \epsilon_{\mu_3 \mu_1 \sigma_1}
 \frac {(x(s_4) - x(s_2))^{\sigma_2}} {|x(s_4)-x(s_2)|^3}
 \frac {(x(s_3) - x(s_1))^{\sigma_1}} {|x(s_3)-x(s_1)|^3}. \nonumber \\
\eea
The total integration range is therefore
\beq
 \Delta_4 := \{ (s_1,s_2,s_3,s_4) |
 1 > s_1 > s_2 > s_3 > s_4 \geq 0 \},
\eeq
and it will be divided into several parts, classified by the location
of the $x(s_i)$ with respect to the crossings.
For every crossing $c$ we define intervals $I_c^+$ and $I_c^-$ in
$\topKreis$ with the following properties (cf. figure \ref{zylinder}):
\beq
x(I_c^+) = g_c^+
\klartext{and}
x(I_c^-) = g_c^- .
\eeq
Due to the construction of the $U_c$, $\pi(0)$ is not contained in any
of the $I_c^{+/-}$ and all intervals $I_c^{+/-}$ are disjoint.
The following cases are considered:
\begin{enumerate}
\item
 $x(s_1),x(s_3) \in U_c\times\RR$,
 with $x(s_1) \in g_c^+$ and $x(s_3) \in g_c^-$,
\newline
 $x(s_2),x(s_4) \in U_d\times\RR$,
 with $x(s_2) \in g_d^+$ and $x(s_4) \in g_d^-$,
\newline
i.e. two pairs of integration variables meet at different crossings
$c$ and $d$ (see figure \ref{sitCD}).
This part will be denoted by $\Delta_4^{(1)}(c,d)$ and is defined as
\beq
 \Delta_4^{(1)}(c,d) = \{ (s_1,s_2,s_3,s_4)\in I_c^+ \times I_d^+
			\times I_c^- \times I_d^-
			\,\, |\,\, s_1 > s_2 > s_3 > s_4 \}.
\eeq
Note that this set can be empty for certain choices of $c$ and $d$.
\item
 $x(s_1),x(s_3) \in U_c\times\RR$,
 with $x(s_1) \in g_c^+$ and $x(s_3) \in g_c^-$,
\newline
 $x(s_2),x(s_4) \in U_c\times\RR$,
 with $x(s_2) \in g_c^+$ and $x(s_4) \in g_c^-$,
\newline
i.e. all integration variables meet at the same crossing $c$ in the way
shown in figure \ref{sitC}. This part will be denoted by $\Delta_4^{(2)}(c)$
and is defined as
\beq
 \Delta_4^{(2)}(c) = \{ (s_1,s_2,s_3,s_4)\in I_c^+ \times I_c^+
			\times I_c^- \times I_c^-
			\,\, |\,\, s_1 > s_2 > s_3 > s_4 \}.
\eeq
\item Other cases, which do not contribute to $\rho_2$. This
part will be called $\Delta_4^{{\rm Rest}}$. These cases contain at least
one pair of integration variables
$(x(s_1),x(s_3))$ or $(x(s_2),x(s_4))$ which do not
encounter at any crossing in the way described in the previous cases.
\end{enumerate}
The total integration range can then be written as
\beq
\Delta_4 =
\bigcup_{c,d \in \kreuzungen} \Delta_4^{(1)}(c,d)
\,\,\,\cup\,\,\,
\bigcup_{c \in \kreuzungen} \Delta_4^{(2)}(c)
\,\,\,\cup\,\,\,
\Delta_4^{{\rm Rest}},
\eeq
and the integral over $K_0$ is written as
\bea
\rho_2(K_0) & =: & \rho_2(K_0;\Delta_4) \nonumber \\
& = &
\sum_{c,d\in\kreuzungen} \rho_2(K_0;\Delta_4^{(1)}(c,d))
+
\sum_{c\in\kreuzungen} \rho_2(K_0;\Delta_4^{(2)}(c))
+
\rho_2(K_0;\Delta_4^{{\rm Rest}}), \nonumber \\
\eea
where the second argument of $\rho_2$ denotes the respective
integration range.
%
%
%
%%%%%%%%%%%%%%%%%%%%%%%
\abbildung{sit_c_d.eps}{\caption{Case 1. The integration variables
encounter at two crossings.} \label{sitCD} }
%%%%%%%%%%%%%%%%%%%%%%%%%%%%%%%%%%%%%%%%%%%%%%%
%%%%%%%%%%%%%%%%%%%%%%%
\abbildung{sit_c.eps}{\caption{Case 2. The integration
 variables encounter at one crossing.} \label{sitC} }
%%%%%%%%%%%%%%%%%%%%%%%%%%%%%%%%%%%%%%%%%%%%%%%
%
%
%
\paragraph{Case 1: Four variables encounter at two crossings.}
The corresponding integration range containing two
crossings $c$ and $d$ is $\Delta_4^{(1)}(c,d)$.
The integral $\rho_2(K_0;\Delta_4^{(1)}(c,d))$ can be split
into two factors:
\bea
\rho_2(K_0;\Delta_4^{(1)}(c,d)) & = &
\frac 1 {8\pi^2}
 \int_{I_c^+} ds_1 \int_{I_c^-} ds_3 \,\,\,
 \dot{x}(s_1)^{\mu_1} \dot{x}(s_3)^{\mu_3}
 \epsilon_{\mu_3 \mu_1 \sigma_1}
 \frac {(x(s_3) - x(s_1))^{\sigma_1}} {|x(s_3)-x(s_1)|^3} \nonumber \\
& & \,\,\,\,\, \times
 \int_{I_d^+} ds_2 \int_{I_d^-} ds_4 \,\,\,
 \dot{x}(s_2)^{\mu_2} \dot{x}(s_4)^{\mu_4}
 \epsilon_{\mu_4 \mu_2 \sigma_2}
 \frac {(x(s_4) - x(s_2))^{\sigma_2}} {|x(s_4)-x(s_2)|^3}. \nonumber \\
\eea
We now calculate one of these factors for the limit of
flat knots. The expression
\beq
\label{einerDerBeidenFaktoren}
 \int_{I_c^+} ds_1 \int_{I_c^-} ds_3 \,\,\,
 \dot{x}(s_1)^{\mu_1} \dot{x}(s_3)^{\mu_3}
 \epsilon_{\mu_3 \mu_1 \sigma_1}
 \frac {(x(s_3) - x(s_1))^{\sigma_1}} {|x(s_3)-x(s_1)|^3}
\eeq
is reparametrizable. We use parameters
$s_1',s_3' \in \ooIntervall{-1}{+1}$ and a corresponding parametrization
$x_1',x_3': \ooIntervall{-1}{+1} \abbilden \RR^3$.
The expression (\ref{einerDerBeidenFaktoren}) is invariant
with respect to scaling, rotating, and translating
the coordinate frame. So the coordinates can be chosen in such a way that
\beq
 x_1'(s_1')^{\mu_1} = (s_1',0,z_1) \mbox{\,\,\, and \,\,\,}
 x_3'(s_3')^{\mu_3} = (0,s_3',z_3)
\eeq
and hence
\beq
 \dot{x}_1'(s_1')^{\mu_1} = (1,0,0) \mbox{\,\,\, and \,\,\,}
 \dot{x}_3'(s_3')^{\mu_3} = (0,1,0).
\eeq
After defining $h=z_1-z_3$ the crossing information of $c$ is simply
\beq
 \epsilon(c) = \sgn(h).
\eeq
The evaluation of the vector products gives
\beq
 \int_{-1}^{+1} ds_1' \int_{-1}^{+1} ds_3' \,\,\,
 \frac h {\sqrt{s_1'^2+s_3'^2+h^2}^{\,3}}.
\eeq
The limit of a flat knot is now equivalent to the limit $h\rightarrow 0$.
In this limit the integral assumes the value
\beq
 2 \pi \epsilon(c),
\eeq
and therefore
\beq
\rho_2(K_0;\Delta_4^{(1)}(c,d)) = \frac 1 2 \epsilon(c) \epsilon(d)
\label{ergebnisCD}
\eeq
if for the crossings $c$ and $d$ the range $\Delta_4^{(1)}(c,d)$ is not
empty.
\paragraph{Case 2: Four variables encounter at one crossing.}
The integration range is
$\Delta_4^{(2)}(c)$.
For this range the integral cannot be factorized as in the previous case.
\bea
\rho_2(K_0;\Delta_4^{(2)}(c)) & = &
 \frac 1 {8\pi^2}
 \int_{I_c^+ \times I_c^+,s_1 > s_2} \mkern -23mu ds_1 \,\, ds_2 \,\,
 \int_{I_c^- \times I_c^-,s_3 > s_4} \mkern -23mu ds_3 \,\, ds_4 \,\,
 \dot{x}(s_1)^{\mu_1}
 \dot{x}(s_2)^{\mu_2}
 \dot{x}(s_3)^{\mu_3}
 \dot{x}(s_4)^{\mu_4} \nonumber \\
& \times &
 \epsilon_{\mu_4 \mu_2 \sigma_2} \epsilon_{\mu_3 \mu_1 \sigma_1}
 \frac {(x(s_4) - x(s_2))^{\sigma_2}} {|x(s_4)-x(s_2)|^3}
 \frac {(x(s_3) - x(s_1))^{\sigma_1}} {|x(s_3)-x(s_1)|^3}. \nonumber \\
\eea
Again, as in case 1, the parametrizations $x_1'$ and $x_3'$ are used,
with parameters in the range $\ooIntervall{-1}{+1}$.
Under the reparametrization we replace
\bea
x(s_1) \rightarrow x_1'(s_1') & \mbox{\,\,\,\,\,and\,\,\,\,\,} &
x(s_2) \rightarrow x_1'(s_2') \nonumber\\
x(s_3) \rightarrow x_3'(s_3') & \mbox{\,\,\,\,\,and\,\,\,\,\,} &
x(s_4) \rightarrow x_3'(s_4') \nonumber\\
\eea
since $s_1$ and $s_2$ relate to the same piece $g_c^+$, and
$s_3$ and $s_4$ relate to $g_c^-$.
We then get
\bea
& & \frac 1 {8\pi^2}
 \int_{-1}^{+1} ds_1' \int_{-1}^{s_1'} ds_2'
 \int_{-1}^{+1} ds_3' \int_{-1}^{s_3'} ds_4' \,\,\,
 \dot{x}_1'(s_1')^{\mu_1}
 \dot{x}_1'(s_2')^{\mu_2}
 \dot{x}_3'(s_3')^{\mu_3}
 \dot{x}_3'(s_4')^{\mu_4} \nonumber \\
 & \times &
 \epsilon_{\mu_4 \mu_2 \sigma_2} \epsilon_{\mu_3 \mu_1 \sigma_1}
 \frac {(x_3'(s_3')-x_1'(s_1'))^{\sigma_1}}
       {|x_3'(s_3')-x_1'(s_1')|^3}
 \frac {(x_3'(s_4')-x_1'(s_2'))^{\sigma_2}}
       {|x_3'(s_4')-x_1'(s_2')|^3} \nonumber \\
\eea
and after evaluating the vector products and using the flat knot limit:
\beq
 \lim_{h\rightarrow 0}
 \frac 1 {8\pi^2}
 \int_{-1}^{+1} ds_1' \int_{-1}^{s_1'} ds_2'
 \int_{-1}^{+1} ds_3' \int_{-1}^{s_3'} ds_4' \,\,\,
 \frac h {\sqrt{s_1'^2+s_3'^2+h^2}^{\,3}}
 \frac h {\sqrt{s_2'^2+s_4'^2+h^2}^{\,3}}
 = \frac 1 8.
\eeq
So the integral $\rho_2(K_0;\Delta_4^{(2)}(c))$ for any crossing
$c\in\kreuzungen$ is
\beq
\rho_2(K_0;\Delta_4^{(2)}(c)) = \frac 1 8.
\label{ergebnisC}
\eeq
\paragraph{Case 3: At least one pair of variables does not encounter
at any crossing.}
We assume that $x(s_1)$ and $x(s_3)$ lie in different parts of the knot,
e.g. in $U\times\RR$ and some $U_c\times\RR$, or in cylinders
associated to different crossings. Then the triple product
\beq
\label{xEinsXDreiFaktor}
 \dot{x}(s_1)^{\mu_1}
 \dot{x}(s_3)^{\mu_3}
 \epsilon_{\mu_3 \mu_1 \sigma_1}
 \frac {(x(s_3) - x(s_1))^{\sigma_1}} {|x(s_3)-x(s_1)|^3}
\eeq
and therefore the whole integral in this range vanishes
in the limit of flat knots.
If $x(s_1)$ and $x(s_3)$ lie in the same cylinder or in $U\times\RR$,
and $x(s_3)$ lies near $x(s_1)$ the expression \gleichungNr{xEinsXDreiFaktor}
vanishes as well, as can be shown by expanding $x(s_3)$ in a power
series near $x(s_1)$.

Using $n$, the self-crossing number $\chi_2$ from section
\ref{crossingNumbers} and the results
\gleichungNr{ergebnisCD} and \gleichungNr{ergebnisC}
$\rho_2(\calK)$ assumes the following form
\beq
\rho_2(\calK) := \lim_{t\abbilden 0} \rho_2(H_t(K_0))
 = \frac 1 2 \chi_2(\calK) + \frac 1 8 n.
\eeq
%
%
%
%%% name: rho1.tex
\section{Construction of $\rho_1(\calK)$ and $\rho^{\rm II}(\calK)$}
\setcounter{equation}{0}
\label{berechnungRhoEins}
The behaviour of $\rho_2(\calK)$ under the Reidemeister moves
follows from the results of section \nolinebreak \ref{verhaltenUnterRM}:
$$
\rho_2(L_{\rm I}^{+/-}) - \rho_2(L_{\rm I}^0) = + \frac 1 8
\,\,\,\,\,\,\,\,\,\,
\rho_2(L_{\rm II-A}) - \rho_2(L_{\rm II-A}^0) = -\frac 1 4
$$
\beq
\rho_2(L_{\rm II-B}) - \rho_2(L_{\rm II-B}^0) = +\frac 1 4
\,\,\,\,\,\,\,\,\,\,
\rho_2(L_{\rm III}^+) - \rho_2(L_{\rm III}^-) = +\frac 1 2.
\eeq
Furthermore, we know that $\rho_2(U_0) = 0$, where $U_0$ is the unknot
diagram without crossing.
We want to construct an ambient isotopy invariant
\beq
\rho^{\rm II}(\calK) = \rho_1(\calK) + \rho_2(\calK),
\eeq
i.e. an object based on the evaluation of $\calK$, invariant under
the Reidemeister moves.
Therefore, we have to postulate a behaviour of $\rho_1$
opposite to that of $\rho_2$:
$$
\rho_1(L_{\rm I}^{+/-}) - \rho_1(L_{\rm I}^0) = - \frac 1 8
\,\,\,\,\,\,\,\,\,\,
\rho_1(L_{\rm II-A}) - \rho_1(L_{\rm II-A}^0) = +\frac 1 4
$$
\beq
\label{RhoEinsBedingungEins}
\rho_1(L_{\rm II-B}) - \rho_1(L_{\rm II-B}^0) = -\frac 1 4
\,\,\,\,\,\,\,\,\,\,
\rho_1(L_{\rm III}^+) - \rho_1(L_{\rm III}^-) = -\frac 1 2.
\eeq
{}From an analytical calculation in \cite{GuMaMi} we know
that
\beq
\label{RhoEinsBedingungZwei}
\rho_1(U_0) = -\frac 1 {12},
\eeq
if $U_0$ represents an unknot that lies within a plane.
In the appendix it will be shown that in the limit of flat knots
$\rho_1$ is independent of the values of the crossing function $\epsilon$.
This means that for any two diagrams $\calK$ and $\calK'$
which differ only in the crossing functions one has
\beq
\rho_1(\calK) = \rho_1(\calK').
\eeq
Therefore, instead of calculating $\rho_1$ for $\calK$ we can use
the standard ascending diagram $\alpha(\calK)$ with respect
to the basepoint:
\beq
\label{RhoEinsBedingungDrei}
\rho_1(\calK) = \rho_1(\alpha(\calK)).
\eeq
For any diagram $\calK$ the standard ascending diagram $\alpha(\calK)$,
as defined in \cite{LiMi},
is obtained by passing through the knot, starting from the basepoint,
and switching each crossing encountered for the first time
to an undercrossing. A standard ascending diagram is a diagram
representation of the unknot.

It should be possible to show directly that the analytical expression
for $\rho_1$ fulfils the above conditions in the limit of flat knots.
Nevertheless,
the properties
(\ref{RhoEinsBedingungEins}),
(\ref{RhoEinsBedingungZwei}), and
(\ref{RhoEinsBedingungDrei})
suffice to construct $\rho_1$ in a well-defined
way and to show that it is unique.
For proving the uniqueness we assume that there is another
expression $\tilde\rho_1$ with the same properties. Let $\calK$ be a knot
diagram. Then $\alpha(\calK)$ can be obtained from $U_0$ by applying
a finite sequence of Reidemeister moves $R_1,\ldots,R_m$:
\beq
\alpha(\calK) = R_m \ldots R_1 \,\, U_0.
\eeq
Since $\tilde\rho_1(U_0) = \rho_1(U_0)$ and the change of both under
Reidemeister moves is the same, also the result is the same.
Hence
\beq
\tilde\rho_1(\calK) = \tilde\rho_1(\alpha(\calK))
=
\rho_1(\alpha(\calK)) = \rho_1(\calK).
\eeq
We now demonstrate that $\rho_1$ is well-defined.
Again, let $\calK$ be a knot which can be obtained from $U_0$
in two different ways, i.e. by applying a sequence $R_1,\ldots,R_m$
or another sequence $R'_1,\ldots,R'_{m'}$.
In principle one has to show that both sequences for $\rho_1$ lead to
the same result. One can, however, simply define
\beq
\rho_1(\calK) = -\frac 1 {12} - \rho_2(\alpha(\calK))
\label{rhoEinsKonstruktion}
\eeq
and check that the defining conditions are fulfilled.
For condition (\ref{RhoEinsBedingungZwei}) this is clear. For
condition (\ref{RhoEinsBedingungDrei}) it is clear as well, since
\beq
\alpha(\alpha(\calK)) = \alpha(\calK).
\eeq
For the properties (\ref{RhoEinsBedingungEins}) some remarks have to be made.
Consider a knot $\calK$ which contains a Reidemeister situation $L$ and
another one $\calK'$ differing from $\calK$ by containing
a Reidemeister situation $L^0$ instead of $L$. Then the move can be
applied to $\calK$. However, if the basepoint is chosen in an
{\sl inappropriate} way,
e.g. between $k_1$ and $k_1+1$ or between $k_2$ and $k_2+1$ in
$L_{\rm II-A}$ of figure \ref{reidemeisterIIA},
the move cannot transform $\alpha(\calK)$ into $\alpha(\calK ')$.
Therefore we will place the basepoint in such a way that the move can be
applied, i.e. {\sl outside} the Reidemeister situation and prove the
independence of $\rho_1$ of the choice of the basepoint afterwards.
If the move can be applied, it is clear by construction that $\rho_1$
behaves correctly.
\paragraph{Proposition.} {\sl $\rho_2(\alpha(\calK))$ is independent of
the choice of the basepoint.}
\paragraph{Proof.} It is sufficient to show that the basepoint can be shifted
by one crossing without changing $\rho_2$. Since the number of crossings
$n$ is invariant under this operation, it is sufficient to show the
invariance of $\chi_2$.

Consider a diagram $\calK=(\pi,\epsilon,\indexmenge)$.
The basepoint $b$ is a point in the piece between two indices
$i_b-1$ and $i_b$. The basepoint shifted by one crossing lies
in the piece between $i_b$ and $i_b+1$.
Let
$\alpha(\calK) = (\pi,\epsilon_\alpha,\indexmenge)$
and
$\alpha'(\calK)= (\pi,\epsilon_{\alpha'},\indexmenge)$
denote the standard ascending
diagrams of $\calK$ with respect to the basepoints $b$ and $b'$.
{}From the definition of the standard ascending diagrams it is
clear that the difference between $\alpha(\calK)$ and $\alpha'(\calK)$
lies in the crossing functions at $i_b$:
\beq
\epsilon_\alpha(i_b) = -\epsilon_{\alpha'}(i_b).
\eeq
The index which forms a crossing together with $i_b$ will be denoted by
$\bar{i_b}\in\indexmenge$.
We have to show that
\beq
\chi_2(\alpha(\calK)) - \chi_2(\alpha'(\calK)) = 0.
\eeq
In section \ref{einKreuzungsVerhalten} it has been shown that
for two diagrams which differ only in one crossing
(in this case $\alpha(\calK)$ and $\alpha'(\calK)$
differ in $\pi(s_{i_b})$), the difference for $\chi_2$ is
\beq
\chi_2(\alpha(\calK)) - \chi_2(\alpha'(\calK)) =
2 \epsilon_\alpha(i_b,\bar{i_b})
\chi_1(\ooIntervall{i_b}{\bar{i_b}},
       \ooIntervall{\bar{i_b}}{i_b};\epsilon_\alpha).
\eeq
The second factor is related to the linking number between
the two link components corresponding to the pieces
$L^1=\pi(\ooIntervall{s_{i_b}}{s_{\bar{i_b}}})$
and $L^0=\pi(\ooIntervall{s_{\bar{i_b}}}{s_{i_b}})$ according to
equation \gleichungNr{relationLinkingChiEins}.
Since $\alpha(\calK)$ is an ascending diagram, the component
$L^1$ lies completely underneath $L^0$, which means that they can
be separated in space. Hence their linking number is zero.
This completes the proof.

Finally, the invariant from the second order term of the
Chern-Simons theory assumes the form
\bea
\rho^{\rm II}(\calK)
& = & \rho_1(\calK) + \rho_2(\calK) \nonumber \\
& = & -\frac 1 {12} + \frac 1 2 \{\chi_2(\calK)-\chi_2(\alpha(\calK))\} .
\label{mainResult}
\eea
This formula is the main result of the present paper.
%
%
%
%%% name: limiconn.tex

\section{Relation between $\rho^{\rm II}$ and the total twist}
\setcounter{equation}{0}
\label{relationRhoIITotalTwisting}
We define
\beq
\tau(\calK) = \frac 1 2 \left( \rho^{\rm II}(\calK)) + \frac {1} {12} \right)
 = \frac 1 4 \{\chi_2(\calK)-\chi_2(\alpha(\calK))\} .
\label{totalTwist}
\eeq
The aim of this section is to show that this is equivalent to the
total twist
which was defined by Lickorish and Millet in \cite{LiMi}.
Consider some knot diagram $\calK = (\pi,\epsilon,\indexmenge)$
and its standard ascending diagram
$\alpha(\calK) = (\pi,\epsilon_\alpha,\indexmenge)$, both of course
with the same
shadow diagram and the same set of crossings $\kreuzungen$.
The difference between the two diagrams may consist in $m$ crossings
$c_1,\ldots ,c_m\in\kreuzungen$ such that
$\epsilon(c_i)=-\epsilon_\alpha(c_i)$ for $i=1\ldots m$.
We shall here use a notation similar to that used in \cite{LiMi}.
Let $\sigma_i , i=1\ldots m$ be the operation applied to a diagram which
switches the crossing $c_i$.
Now define $\calK_0 = \alpha(\calK)$ and
$\calK_j = (\pi,\epsilon_j,\indexmenge)$ with
\beq
\calK_j = \sigma_j \sigma_{j-1} \ldots \sigma_1 \calK_0
\eeq
so that $\calK = \calK_m$.
Then $\tau$ can be written as
\beq
\tau(\calK) = \frac 1 4 \{\chi_2(\calK_m)-\chi_2(\calK_0)\}.
\eeq
For every $\calK_j$ and every $c\in\kreuzungen$ formed by the
indices $i_c^+$ and $i_c^-$ we define the non-overlapping pieces
$S_j^+(c) = \pi(\ooIntervall{s_{i_c^+}}{s_{i_c^-}})$ and
$S_j^-(c) = \pi(\ooIntervall{s_{i_c^-}}{s_{i_c^+}})$ and the link components
$L_j^+(c)$ and $L_j^-(c)$,
which arise from nullifying the crossing $c$.
The two slightly different notations are used to draw the connection
between our formulation and the one in \cite{LiMi}.
In section \ref{einKreuzungsVerhalten} we have already calculated
the change of $\chi_2$ under the change of one crossing, or in this case
the change under the operation $\sigma_j$.
The result can be written as
\beq
\chi_2(\calK_{j})-\chi_2(\calK_{j-1})
=
2 \epsilon_j(c_j) \chi_1(S_j^+(c_j),S_j^-(c_j)).
\label{mirFaelltNixEin}
\eeq
Using relation \gleichungNr{mirFaelltNixEin} we write $\tau(\calK)$ as
\bea
\tau(\calK) & = & \frac 1 4
	\left\{
	  \chi_2(\calK_m)-\chi_2(\calK_{m-1})
	+ \chi_2(\calK_{m-1})-\chi_2(\calK_{m-2})
	+\ldots
	+\chi_2(\calK_1)-\chi_2(\calK_0)
	\right\} \nonumber \\
& = & \frac 1 2 \sum_{j=1}^m \epsilon_j(c_j) \chi_1(S_j^+(c_j),S_j^-(c_j)).
\eea
The link components $L_j^+(c_j)$ and $L_j^-(c_j)$,
which correspond to the pieces $S_j^+(c_j)$ and $S_j^-(c_j)$ in the
previous equation, coincide precisely
with the definitions of $L^1_j$ and $L^0_j$ used in \cite{LiMi}.
Therefore, using $\epsilon_j(c_j) = \epsilon(c_j)$, the definitions
of $L^1_j$ and $L^0_j$, and the relation
\gleichungNr{relationLinkingChiEins} between $\lambda$ and $\chi_1$ one
obtains
\beq
\tau(\calK) = \sum_{j=1}^m \epsilon(c_j) \lambda(L^1_j,L^0_j),
\eeq
which is the same expression as the one given in \cite{LiMi}.
%
%
%
%%% name:example.tex
\section{Example: Calculation of $\rho^{\rm II}$}
\setcounter{equation}{0}
\label{beispiel}
Finally, we shall demonstrate the calculation of $\rho^{\rm II}$
for a specific knot. We choose the knot $5_2$ in the notation
of Rolfsen in \cite{Rolfsen}, shown below in figure \ref{knotenFuenfZwei}.
The {\sl calculation} consists of
listing all possible non-vanishing contributions in the fourfold
sums $\chi_2$ with crossing function $\epsilon$
and its standard ascending version with crossing function
$\epsilon_\alpha$. In table 1 below all significant combinations
of indices for two variables $j_1$, $j_3$ or $j_2$, $j_4$ fulfilling
$j_1>j_3$ and $j_2>j_4$ are listed, together with the corresponding values
of $\epsilon$ and $\epsilon_\alpha$.
{}From this table all non-vanishing contributions for $j_1>j_2>j_3>j_4$
are constructed (see table 2).
The results are
\beq
\chi_2(5_2) = +7 \mbox{\,\,\,\,\, and \,\,\,\,\,} \chi_2(\alpha(5_2)) = -1,
\eeq
and therefore
\beq
\rho^{\rm II}(5_2) = \frac {47} {12}
\mbox{\,\,\,\,\, and \,\,\,\,\,}
\tau(5_2) = 2.
\eeq
%%%%%%%%%%%%%%%%%%%%%%%%%%%%%%%%%%%
\abbildung{knot_5_2.eps}{\caption{The knot $5_2$
with basepoint and its standard ascending diagram.}
\label{knotenFuenfZwei} }
%%%%%%%%%%%%%%%%%%%%%%%%%%%%%%%%%%%%%%%%%%%%%%
\begin{table}[h]
\centering
\begin{tabular}{cccc}
$j_1$ & $j_3$ & $\epsilon(j_1,j_3)$ & $\epsilon_\alpha(j_1,j_3)$ \\
$j_2$ & $j_4$ & $\epsilon(j_2,j_4)$ & $\epsilon_\alpha(j_2,j_4)$ \\
\hline
5 & 0 & $-$ & $+$ \\
6 & 1 & $-$ & $-$ \\
7 & 4 & $-$ & $+$ \\
8 & 3 & $-$ & $-$ \\
9 & 2 & $-$ & $+$ \\
\label{koinzidenztabelleZwei}
\end{tabular}
\caption{Encounters of two variables in the knot $5_2$.}
\end{table}
\begin{table}[ht]
\centering
\begin{tabular}{cccccc}
$j_1$ & $j_2$ & $j_3$ & $j_4$ &
$\epsilon(j_1,j_3)\epsilon(j_2,j_4)$ &
$\epsilon_\alpha(j_1,j_3)\epsilon_\alpha(j_2,j_4)$ \\
\hline
6 & 5 & 1 & 0 & $+$ & $-$ \\
7 & 5 & 4 & 0 & $+$ & $+$ \\
7 & 6 & 4 & 1 & $+$ & $-$ \\
8 & 5 & 3 & 0 & $+$ & $-$ \\
8 & 6 & 3 & 1 & $+$ & $+$ \\
9 & 5 & 2 & 0 & $+$ & $+$ \\
9 & 6 & 2 & 1 & $+$ & $-$ \\
\hline
  &   &   &   & $+7$ & $-1$ \\
\label{koinzidenztabelleVier}
\end{tabular}
\caption{All non-vanishing contributions to
$\chi_2(\calK)$ and $\chi_2(\alpha(\calK))$.}
\end{table}
\pagebreak
\section{Outlook}
\setcounter{equation}{0}
As already emphasized, the method of flattening the knot in order to
calculate the complicated line integrals from the perturbative expansion
of the Wilson loops can be applied to higher orders as well.
We have already checked this by the use of a computer program in
C++ to be run on a PC which calculates
$\rho^{\rm II}(\calK)$ and the third order invariant
$\rho^{\rm III}(\calK)$ for arbitrary knots.
Whoever is interested in this program may order it via e-mail.
We shall soon publish a discussion of the third order calculation
which employs the formalism developed in this paper.

One may hope that a systematic examination of the higher orders
will yield a convenient description of the Vassiliev invariants, which
are conjectured to classify the knots uniquely
\cite{Va}.
The computation time for the invariants $\rho^{\rm II}$ and
$\rho^{\rm III}$ using the procedure presented here grows as
$O(n^2)$ and $O(n^3)$ respectively. This is to be compared to computations
involving polynomial invariants, whose complexity grows exponentially
with crossing number \cite{BiLi}.
\newline
\newline
\noindent {\bf Acknowledgements.}
It is a pleasure to thank E. Guadagnini and M. Mintchev
for many helpful discussions, and for their warm hospitality
during our visit in Pisa.
\begin{appendix}
%
%%% rho1inde.tex

\section*{Appendix\newline Independence of $\rho_1$ with respect to the
values of $\epsilon$}
\setcounter{equation}{0}
\renewcommand{\theequation}{A.\arabic{equation}}
It will now be demonstrated that $\rho_1(\calK)$ is not dependent
on the values of the crossing information $\epsilon$, i.e. that $\rho_1$ can
be interpreted as a functional of the shadow diagram.
The analytical form of $\rho_1$ for a knot $K_0$ is given by
\cite{GuMaMi}
\bea
\rho_1(K_0) & = &
-\frac 1 {32\pi^3}
    \int_{K_0} dx_1^{\mu_1}
    \int_{\rm BP}^{x_1} dx_2^{\mu_2}
    \int_{\rm BP}^{x_2} dx_3^{\mu_3} \,\,
\epsilon^{\nu_1 \nu_2 \nu_3}
\epsilon_{\mu_1 \nu_1 \sigma_1}
\epsilon_{\mu_2 \nu_2 \sigma_2}
\epsilon_{\mu_3 \nu_3 \sigma_3}
I^{\sigma_1 \sigma_2 \sigma_3}(x_1,x_2,x_3) \nonumber \\
\eea
with
\beq
I^{\sigma_1 \sigma_2 \sigma_3}(x_1,x_2,x_3)
=
\int d^3z \frac {(z-x_1)^{\sigma_1}} {|z-x_1|^3}
          \frac {(z-x_2)^{\sigma_2}} {|z-x_2|^3}
          \frac {(z-x_3)^{\sigma_3}} {|z-x_3|^3}.
\eeq
It was also demonstrated in \cite{GuMaMi}
that this integral can be solved and assumes the form
\beq
I^{\sigma_1 \sigma_2 \sigma_3}(x_1,x_2,x_3)
=
-\frac {\partial} {\partial x_2^{\sigma_2}}
 \frac {\partial} {\partial x_3^{\sigma_3}}
I^{\sigma_1}(x_2-x_1,x_1-x_3)
\label{ddIForm}
\eeq
with
\beq
I^{\sigma_1}(c,b)
=
2 \pi \frac {|c|+|b|-|c+b|} {|c||b|-c\cdot b}
\left\{
  \frac {c^{\sigma_1}} {|c|}
  -
  \frac {b^{\sigma_1}} {|b|}
\right\}
\eeq
where a slightly different notation has been used in comparison to
\cite{GuMaMi}
in order to formulate the expression in a symmetric way.
If we define
\beq
a = x_3-x_2,
\,\,\,\,\,
b = x_1-x_3,
\,\,\,\,\,
c = x_2-x_1
\eeq
so that $a=-b-c$, the expression $I^{\sigma_1}$ can be written as
\bea
I^{\sigma_1}(b,c)
& = &
4 \pi \frac 1 {|a|+|b|+|c|}
\left\{
  \frac {c^{\sigma_1}} {|c|}
  -
  \frac {b^{\sigma_1}} {|b|}
\right\} \nonumber \\
& = &
\left\{
   \frac {\partial} {\partial c^{\sigma_1}}
  -\frac {\partial} {\partial b^{\sigma_1}}
\right\}
4\pi\ln (|a|+|b|+|c|). \nonumber \\
\eea
$a$, $b$, $c$ will be considered as functions of $x_1,x_2,x_3$
in the following.
The partial derivatives in (\ref{ddIForm})
can be reformulated as partial derivatives with respect to
$a$, $b$, and $c$.
Then $I^{\sigma_1 \sigma_2 \sigma_3}$ is formulated in a symmetric way:
\beq
I^{\sigma_1 \sigma_2 \sigma_3}(x_1,x_2,x_3)
=
\left\{
   \frac {\partial} {\partial b^{\sigma_1}}
  -\frac {\partial} {\partial c^{\sigma_1}}
\right\}
\left\{
   \frac {\partial} {\partial c^{\sigma_2}}
  -\frac {\partial} {\partial a^{\sigma_2}}
\right\}
\left\{
   \frac {\partial} {\partial a^{\sigma_3}}
  -\frac {\partial} {\partial b^{\sigma_3}}
\right\}
4\pi\ln (|a|+|b|+|c|).
\eeq
This can be calculated straightforwardly.
We use the abbreviations
\beq
N = |a|+|b|+|c|,
\,\,\,\,\,
m^{\sigma}(u,v)
=
\frac {u^{\sigma}} {|u|}
-
\frac {v^{\sigma}} {|v|},
\,\,\,\,\,
m^{\sigma \tau}(u)
=
\frac 1 {|u|} \delta^{\sigma \tau}
-
\frac {u^{\sigma} u^{\tau}} {|u|^3}
\eeq
and obtain
\bea
I^{\sigma_1 \sigma_2 \sigma_3}(x_1,x_2,x_3)
& = &
4 \pi
\left(
  \frac 2 {N^3} m^{\sigma_1}(b,c)m^{\sigma_2}(c,a)m^{\sigma_3}(a,b)
\right. \nonumber \\
& + &
\left.
  \frac 1 {N^2}
  \left(
     m^{\sigma_1\sigma_3}(b) m^{\sigma_2}(c,a)
    +m^{\sigma_3\sigma_2}(a) m^{\sigma_1}(b,c)
    +m^{\sigma_2\sigma_1}(c) m^{\sigma_3}(a,b)
  \right)
\right). \nonumber \\
\eea
Using a parametrization $x(s)$ and the abbreviations $x_i := x(s_i)$,
$\dot{x}_i := \dot{x}(s_i)$, $\hat{u}$ for the unit vector of any vector $u$,
the line integral $\rho_1$ assumes the form
\bea
\rho_1(K_0)
 =
\frac 1 {8\pi^2}
\int_0^1 ds_1\,\,
\int_0^{s_1} ds_2\,\,
\int_0^{s_2} ds_3
\!\!\!\!\!\!\!\!
& &
\frac 2 {N^3} \,
[\dot{x}_1 \times m(b,c)] \cdot
[\dot{x}_2 \times m(c,a)] \times
[\dot{x}_3 \times m(a,b)] \nonumber \\
& + &
\frac 1 {N^2}
\left\{
  \dot{x}_1 \times m(b,c) \cdot
  \frac 1 {|a|}
  (
      \dot{x}_2 \times \dot{x}_3
    - \hat{a}(\dot{x}_2 \times \dot{x}_3 \cdot \hat{a})
  )
\right. \nonumber \\
& & \,\,\,\,\,\, + \,\,
\left.
\dot{x}_2 \times m(c,a) \cdot
\frac 1 {|b|}
  (
      \dot{x}_3 \times \dot{x}_1
    - \hat{b}(\dot{x}_3 \times \dot{x}_1 \cdot \hat{b})
  )
\right. \nonumber \\
& & \,\,\,\,\,\, + \,\,
\left.
  \dot{x}_3 \times m(a,b) \cdot
  \frac 1 {|c|}
  (
      \dot{x}_1 \times \dot{x}_2
    - \hat{c}(\dot{x}_1 \times \dot{x}_2 \cdot \hat{c})
  )
\right\}. \nonumber \\
\eea
Some properties of the integrand will now be examined.
We shall refer to it as $h(x_1,x_2,x_3) = h(x(s_1),x(s_2),x(s_3))$ so that
\beq
\rho_1(K_0;\Delta_3) := \rho_1(K_0) = \int_{\Delta_3}
 ds_1\,\,ds_2\,\,ds_3\,\,\,h(x(s_1),x(s_2),x(s_3)),
\eeq
where
\beq
\Delta_3 = \{ (s_1,s_2,s_3) \mid 1>s_1>s_2>s_3\geq0 \}.
\eeq
The property of $h(x_1,x_2,x_3)$ which is important here is its
invariance under parity transformations.
As a consequence
\beq
\label{beidseitigerLimes}
(\lim_{t\rightarrow 0 \atop t>0} - \lim_{t\rightarrow 0 \atop t<0})
h(H_t \circ x(s_1),
  H_t \circ x(s_2),
  H_t \circ x(s_3)) = 0,
\eeq
where the function $H_t$ from section \ref{limesFlacherKnoten}
has been used, now however with $t\in\RR$.
The integration range can be split into three subsets.
\begin{enumerate}
\item
$x(s_1),x(s_2),x(s_3) \in U_c\times\RR $ and at least one of the
$x(s_i)$ is in
$g_c^+$ and another in $g_c^-$.
This part will be denoted as $\Delta_3^{(1)}(c)$.
\item
$x(s_2),x(s_3) \in U_c\times\RR $, $x(s_1) \not\in U_c\times\RR $, with
$x(s_2) \in g_c^+$ and $ x(s_3) \in g_c^-$, and
further permutations. These parts together will be denoted as
$\Delta_3^{(2)}(c)$.
\item
Other cases, where no two variables cross each other.
These situations are contained in $\Delta_3^{{\rm Rest}}$.
\end{enumerate}
Consider now case 1.
We take a knot diagram $\calK = (\pi,\epsilon,\indexmenge)$
as flat knot limit of a knot $K$
and another diagram $\calK'=(\pi,\epsilon',\indexmenge)$
as limit of $K'$
with the sole difference that $\epsilon(c) = -\epsilon'(c)$ at a
crossing $c\in\kreuzungen$.
Consider now the cylinder $U_c\times\RR$.
Within this cylinder the flat knot limits of the
knots $K$ and $K'$ differ only by a parity
transformation.

Hence the function $h$ with arguments
in $U_c\times\RR$ is the same for $K$ and $K'$ and therefore independent
of $\epsilon(c)$ and $\epsilon'(c)$ in the range $\Delta_3^{(1)}(c)$.

For case 2 with $x(s_2)$ and $x(s_3)$ encountering each other
at $c$ we consider equation (\ref{beidseitigerLimes}). Since $h$ is a
continuous function in all three arguments the expression can be
replaced by
\beq
(\lim_{t\rightarrow 0 \atop t>0} - \lim_{t\rightarrow 0 \atop t<0})
h(H_0 \circ x(s_1),
  H_t \circ x(s_2),
  H_t \circ x(s_3)) = 0.
\eeq
The two limits correspond to flattening $K$ to $\calK$ and $K'$ to $\calK'$.
Since the last expression vanishes, the function $h$ is independent of
$\epsilon(c)$ in the range $\Delta_3^{(2)}(c)$.

We now consider case 3. In the range $\Delta_3^{{\rm Rest}}$ the function
$
h(H_t \circ x(s_1),
  H_t \circ x(s_2),
  H_t \circ x(s_3))
$
is well defined for $t=0$ because no two variables coincide at any
crossing.
The independence of $h$ on the values of the crossing function is clear,
because at every variable both limits $t\rightarrow 0$ lead to
the same result, and because of the continuity of $h$ one can set $t=0$.
\end{appendix}
%
%%% name: rho_bib.tex

%
\end{document}